\documentclass[prd,aps,twocolumn,a4paper,floatfix,showpacs,nofootinbib]{revtex4-1}

\usepackage[utf8]{inputenc}  
\usepackage[T1]{fontenc}
\usepackage{graphicx,psfrag}
\usepackage{mathrsfs}
\usepackage{amsmath,amsfonts,amssymb,gensymb}
\usepackage{multirow,enumerate}
\usepackage{comment}
\usepackage[hidelinks]{hyperref}
\usepackage{color}
\usepackage{xcolor}
\usepackage{acronym}
\usepackage{xspace}
\usepackage[normalem]{ulem}
\usepackage{mathtools}
\usepackage{subfigure}
\usepackage{makecell}
\usepackage{appendix}

\newacro{BH}{black hole}
\newacro{NS}{neutron star}
\newacro{PN}{Post-Newtonian}
\newacro{BBH}{binary black hole}
\newacro{BNS}{binary neutron star}
\newacro{EOB}{effective-one-body}
\newacro{NR}{numerical relativity}
\newacro{GW}{gravitational wave}
\newacro{EOS}{equation-of-state}

\newcommand{\be}{\begin{equation}}
\newcommand{\ee}{\end{equation}}
\newcommand{\bea}{\begin{eqnarray}}
\newcommand{\eea}{\end{eqnarray}}
\newcommand{\bel}{\begin{align}}
\newcommand{\eel}{\end{align}}

\def\GMc2{{\rm G M_{\odot} c^{-2}}}

\def\SEOBNRv4T{\texttt{SEOBNRv4T}\xspace}

\usepackage{color}
\definecolor{cyan}{rgb}{0,0.9,0.9}
\definecolor{orange}{rgb}{0.9,0.5,0}
\definecolor{magenta}{rgb}{1,0,1}
\definecolor{purple}{rgb}{0.8,0.4,0.8}
\definecolor{gray}{rgb}{0.5,0.5,0.5}
\definecolor{mygreen}{rgb}{0.1,0.8,0.1}
\definecolor{darkblue}{rgb}{0.0,0.0,0.6}

\begin{document}

\title{Measuring tidal effects with the Einstein Telescope: A design study} 

\author{Anna Puecher$^{1,2}$}
\author{Anuradha Samajdar$^{3}$}
\email{anuradha.samajdar@uni-potsdam.de}
\author{Tim Dietrich$^{3,4}$}

\affiliation{${}^1$Nikhef -- National Institute for Subatomic Physics, 
	Science Park 105, 1098 XG Amsterdam, The Netherlands}
\affiliation{${}^2$Institute for Gravitational and Subatomic Physics (GRASP), 
	Utrecht University, Princetonplein 1, 3584 CC Utrecht, The Netherlands}
\affiliation{${}^3$Institut f\"{u}r Physik und Astronomie, Universit\"{a}t Potsdam, Haus 28, Karl-Liebknecht-Str. 24/25, 14476, Potsdam, Germany}
\affiliation{${}^4$Max Planck Institute for Gravitational Physics (Albert Einstein Institute), Am M\"uhlenberg 1, Potsdam 14476, Germany}

\date{\today}

\begin{abstract}
Over the last few years, there has been a large momentum to ensure that the third-generation era of gravitational wave detectors will find its realisation in the next decades, and numerous design studies have been ongoing for some time. 
Some of the main factors determining the cost of the Einstein Telescope lie in the length of the interferometer arms and its shape: L-shaped detectors versus a single triangular configuration. Both designs are further expected to include a xylophone configuration for improvement on both ends of the frequency bandwidth of the detector. 
We consider binary neutron star sources in our study, as examples of sources already observed with the current generation detectors and ones which hold most promise given the broader frequency band and higher sensitivity of the third-generation detectors. We estimate parameters of the sources, with different kinds of configurations of the Einstein Telescope detector, varying arm-lengths as well as shapes and alignments. Overall, we find little improvement with respect to changing the shape, or alignment. However, there are noticeable differences in the estimates of some parameters, including tidal deformability, when varying the arm-length of the detectors. In addition, we also study the effect of changing the laser power, and the lower limit of the frequency band in which we perform the analysis. 
\end{abstract}

\maketitle

\section{Introduction}
\label{sec:intro}
The observation of the first gravitational-wave (GW) signal from a binary neutron star (BNS) source, GW170817~\cite{LIGOScientific:2017vwq} by the LIGO-Scientific~\cite{TheLIGOScientific:2014jea} and Virgo~\cite{TheVirgo:2014hva} Collaborations, has 
provided a wealth of information, starting from constraining the expansion rate of the Universe~\cite{Abbott:2017xzu,Bulla:2022ppy}, to establishing neutron star mergers as one of the main cosmic sources of \emph{r-}process 
elements~\cite{Eichler:1989ve,Rosswog:1998hy,Cowperthwaite:2017dyu,Smartt:2017fuw,Kasliwal:2017ngb,Kasen:2017sxr,Tanvir:2017pws,Rosswog:2017sdn,Abbott:2017wuw,Ascenzi:2018mbh,Watson:2019xjv}, allowing a stringent bound on the speed of GWs~\cite{GBM:2017lvd}, and placing constraints on alternative theories of gravity~\cite{Ezquiaga:2017ekz,Baker:2017hug,Creminelli:2017sry,Sakstein:2017xjx}. The supranuclear equation-of-state (EOS) was constrained from the observation of GW170817~\cite{LIGOScientific:2018hze,LIGOScientific:2018cki}. 
Furthermore, the simultaneous observation of electromagnetic (EM) radiation observed with GW170817~\cite{LIGOScientific:2017ync, LIGOScientific:2017zic, LIGOScientific:2017pwl} helped bounding the supranuclear EOS of neutron stars even more stringently~\cite{Bauswein:2017vtn,Ruiz:2017due,Radice:2017lry,Most:2018hfd,Coughlin:2018fis,Capano:2019eae,Dietrich:2020efo,Breschi:2021tbm,Nicholl:2021rcr,Raaijmakers:2021slr,Huth:2021bsp}. 

\begin{figure}[t]
	\centering
	\includegraphics[width=0.5\textwidth,height=0.5\textwidth]{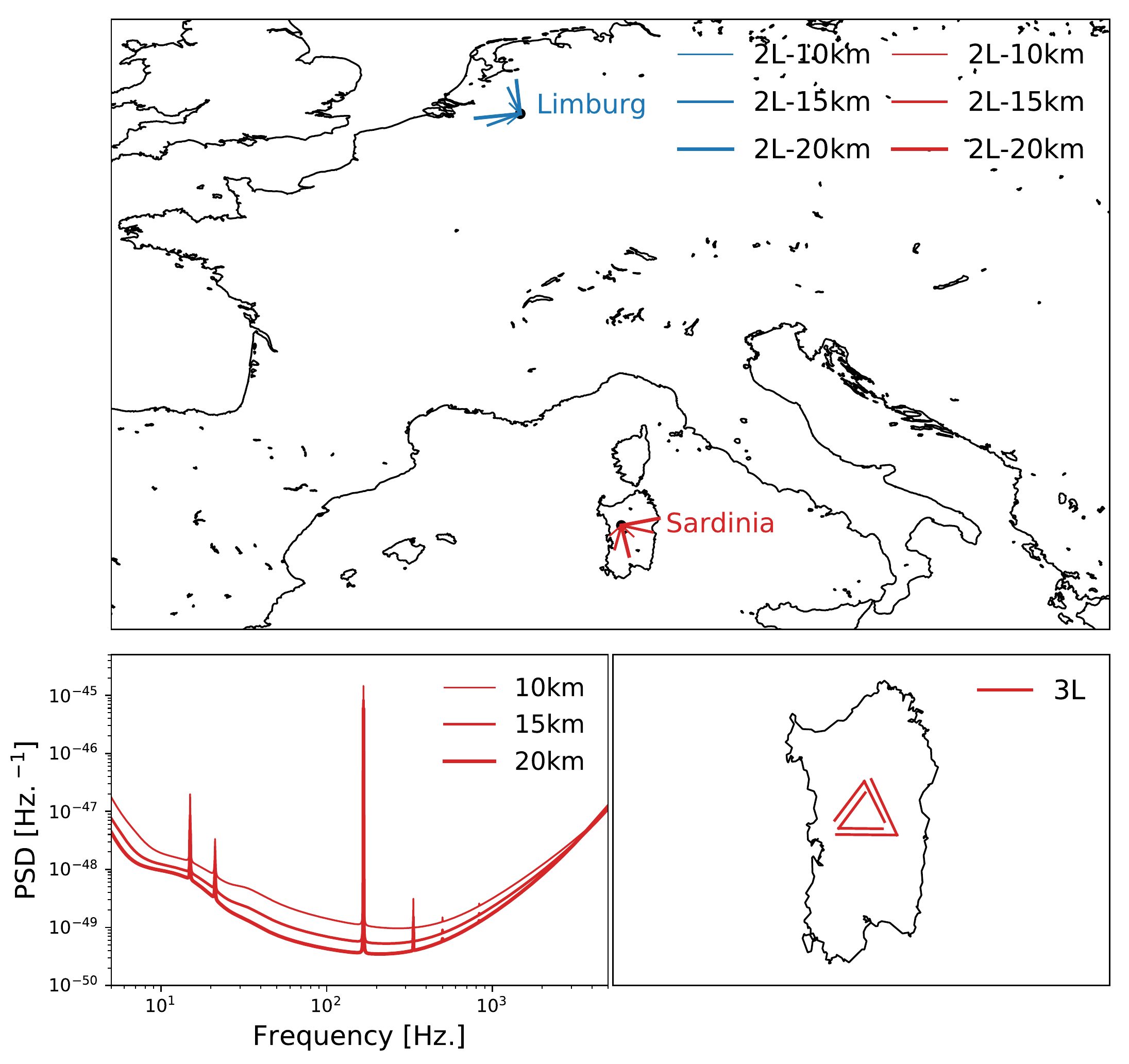}
	\caption{ \textit{Top panel:} Location of the ET detectors in case of two L-shaped interferometers, showing also the possible different arm-lengths in misaligned orientations. \textit{Bottom panel (left):} Power spectral density (PSD) curves for the xylophone configuration in cryogenic mode for different arm-lengths. \textit{Bottom panel (right):} Representation of the ET triangular configuration, located in Sardinia for the purpose of this work.
	}
	\label{fig:et_config}
\end{figure}

As a detector with a much wider frequency band sensitive to GWs, the Einstein Telescope (ET)~\cite{Freise:2008dk,Hild:2009ns,Punturo:2010zz,Sathyaprakash:2011bh,Maggiore:2019uih} promises to observe BNS signals for many cycles, increasing the detected signals' duration up to an hour. The analog detector expected to be operational in the US is Cosmic Explorer (CE) \cite{Evans:2021gyd,Reitze:2019iox}. ET will certainly provide more constrained bounds on the neutron star EOS, even without an accompanying EM counterpart \cite{Pacilio:2021jmq, Maselli:2020uol, Sabatucci:2022qyi, Iacovelli:2022bbs, Williams:2022vct, Breschi:2021xrx, Rose:2023uui}. 

Unfortunately, it is very challenging to perform realistic studies exploiting the full capability of ET due to the wide frequency range it will cover and the large associated computational costs, however, numerous progress has been made regarding GW searches for ET signals, e.g.,\cite{Meacher:2015rex,Wu:2022pyg}, and full parameter estimation (PE) studies, as discussed below.
Ref.~\cite{Smith:2021bqc} performed the first full Bayesian estimation study of GW signals from BNS sources observed by ET with a lower frequency cutoff $f_{\rm{low}}$ of 5 Hz. For this, they constructed reduced order quadratures to make their study computationally feasible.  
Here, we perform PE using \emph{relative binning}~\cite{Zackay:2018qdy, Dai:2018dca, Leslie:2021ssu} to reduce the computational cost of our analysis.
We use a lower frequency of $f_{\mathrm{low}} = 6 \rm Hz$, to provide realistic estimates of the source parameters, such as the chirp mass or the tidal deformability.\\
In the past, various PE studies have already been performed to estimate the Science returns to constrain tidal deformability from BNS observations in ET, including using ET in a network of detectors e.g.,~\cite{Pacilio:2021jmq,Maselli:2020uol,Sabatucci:2022qyi,Iacovelli:2022bbs,Williams:2022vct,Breschi:2021xrx,Sathyaprakash:2009xt,Smith:2021bqc,VanDenBroeck:2010vx,Zhao:2010sz,Punturo:2010zza,Sathyaprakash:2011bh,Sathyaprakash:2012jk,Broeck:2013rka,Maggiore:2019uih,Sathyaprakash:2019yqt,Samajdar:2020xrd,Pacilio:2021jmq,Nitz:2021pbr,Chan:2018csa,Zhao:2017cbb,HernandezVivanco:2019vvk,Castro:2022mpw,Puecher:2022oiz,Rose:2023uui,Sabatucci:2022qyi}. However, up to our knowledge, all of them have focused on the originally proposed triangular design with three interferometers, each having a 60\degree opening angle and arm-length of 10 km, arranged in an equilateral triangle. 

Recently, there has been an increasing interest in studying also different detector configurations and layouts. With this regard, Ref.~\cite{Branchesi:2023mws} provided a detailed discussion with respect to numerous scientific cases and how they are affected by different proposed designs for ET, including different arm-lengths and shapes. More explicitly, Ref.~\cite{Branchesi:2023mws} considered the originally conceived triangular configuration, as well as two separate L-shaped detectors; for the latter, also different alignments, i.e., orientation between the detectors.
In this study, we compare these different designs, cf. Fig.~\ref{fig:et_config}, for recovery of tidal deformability and the other parameters of BNS systems  via full PE analysis. 
In all the studied cases, we analyze BNS simulations, and look at differing results from varying the detector setup, keeping the source properties of the BNS system and the settings of the PE analyses unchanged. Given the significant costs of ET, our study may provide useful information to make an estimate based on the Science returns to reach a more informed decision about the final configuration of the detector. Some of our results have already been presented in \cite{Branchesi:2023mws}, but in a much more compressed and less detailed form. \\
We provide details of the PE methods in Sec.~\ref{sec:method} and give the results of our study in Sec.~\ref{sec:results}. We provide a summary and conclude in Sec.~\ref{sec:conclusion}.

\section{Methods}
\label{sec:method}
Following standard techniques, we use a Bayesian analysis to construct posterior probability density functions (PDFs) on the parameters of interest, i.e., those characterizing the GW waveform describing a BNS merger. As we use quite low values of $f_{\rm low}$ to make our estimates realistic with what is envisaged for the ET detector, our likelihood integral calculation is computationally expensive. For this reason, we resort to the technique of \emph{relative binning} following the analyses of GW170817 in~\cite{Zackay:2018qdy,Dai:2018dca,Leslie:2021ssu}. This approach reduces our computational costs noticeably and makes our runs computationally feasible.

\subsection{Bayesian analysis}
In the employed Bayesian framework, all information about the parameters 
of interest is encoded in the PDF, given 
by Bayes' theorem \cite{Veitch:2009hd}:
\begin{equation}
 p(\vec{\theta}|\mathcal{H}_s,d) 
 = \frac{p(d|\vec{\theta},\mathcal{H}_s)\,p(\vec{\theta}|\mathcal{H}_s)}{p(d|\mathcal{H}_s)},
 \label{eqn:Bayes}
\end{equation}
where $\vec{\theta}$ is the set of parameter values and $\mathcal{H}_s$ is the hypothesis that a GW signal depending on the parameters $\vec{\theta}$ is present in the data $d$. 
For parameter estimation purposes, the factor $p(d|\mathcal{H}_s)$, called the \emph{evidence} for the hypothesis  $\mathcal{H}_s$, is effectively set by the requirement that PDFs are normalized. Assuming the noise to be Gaussian, the \emph{likelihood} $p(d|\vec{\theta},\mathcal{H}_s)$ of obtaining data $d(f)$ given the 
presence of a signal $h(f)$ is determined by the proportionality
\begin{equation}
 p(d|\vec{\theta},\mathcal{H}_s) \propto \exp{\left [-\frac{1}{2}(d-h|d-h)\right ]},
 \label{eqn:lhood}
\end{equation}
where the noise-weighted inner product $(\,\cdot\,|\,\cdot\,)$ is defined as
\begin{equation}
(d|h) = 4\Re \int_{f_{\rm low}}^{f_{\rm high}} \frac{\tilde{d}(f)\,\tilde{h}^\ast(f)}{S_h(f)}\,df.
\label{eqn:inner_product}
\end{equation}
Here a tilde refers to the Fourier transform, and $S_h(f)$ is the power spectral density (PSD). 

Our choices for the \emph{prior probability density} $p(\vec{\theta}|\mathcal{H}_s)$ in Eq.~(\ref{eqn:Bayes}) 
are similar to what has been used for the analyses of real data when BNS signals were 
present with masses similar to GW170817. 
To sample the likelihood function in Eq.~(\ref{eqn:lhood}), we use the \texttt{Bilby} library 
\cite{Ashton:2018jfp,Romero-Shaw:2020owr}, and specifically \texttt{dynesty}~\cite{Speagle:2019ivv,sergey_koposov_2022_6456387} algorithm. 
The waveform we use for both signal injection and recovery is \texttt{IMRPhenomD\_NRTidalv2} 
\cite{Dietrich:2017aum,Dietrich:2018uni,Dietrich:2019kaq}.

\subsection{Relative Binning}
The likelihood integral shown in Eq.~\eqref{eqn:lhood} is constructed over a grid of frequencies and becomes computationally expensive as both the range of the integral grows, and as we analyze longer waveforms like those of BNS sources, which last for many cycles in the frequency band.\\
For our PE studies, we have varied $f_{\rm{low}}$ starting from 6~Hz up to 20~Hz, making the duration of the BNS signal in band from about 75 to about 3 minutes, respectively. In addition, for an inference study to estimate parameters, we generate  millions of waveforms, each associated with its own likelihood value for the sampling to get updated and for points to move towards higher likelihood regions. 
The method of relative binning~\cite{Zackay:2018qdy,Dai:2018dca,Leslie:2021ssu} restricts the number of waveform evaluations by computing the likelihood from summary data calculated on a dense frequency grid for only one \emph{fiducial} waveform, which must resemble the best fit to the data. The underlying assumption is that the set of parameters yielding a non-negligible contribution to the posterior probability produce similar waveforms, such that their ratio varies smoothly in	the frequency domain. In this case, within each frequency bin $b = [f_{\mathrm{min}}(b), f_{\mathrm{max}}(b)]$, the ratio between the sampled waveforms and the fiducial one can be approximated by a linear function in frequency
\begin{equation}
r(f) = \frac{h(f)}{h_0(f)} = r_0(h, b) + r_1(h, b)(f - f_m(b)) + \cdots,
\label{eqn:ratio}
\end{equation}
where $h_0$ is the fiducial waveform and $f_m(b)$ the central frequency of the frequency bin $b$.
 
The coefficients $r_0(h,b)$ and $r_1(h,b)$ are computed for each sampled waveform, but can be determined from the values of $r(f)$ at the edges of the frequency bin $b$. 
Eq.~\ref{eqn:inner_product} can be written in terms of Eq.~\ref{eqn:ratio} and in a discrete form as
\begin{equation}
 (d|h) \approx \sum_b \left( A_0(b) r_0^\ast(h,b) + A_1(b) r_1^\ast(h,b) \right),
\end{equation}
where the summary data
\begin{align}
A_0(b) &= 4 \sum_{f \in b} \frac{d(f) h_0^*(f)}{S_n(f)/T}, \\
A_1(b) &= 4 \sum_{f \in b} \frac{d(f) h_0^*(f)}{S_n(f)/T} (f-f_m(b)) 
\end{align}
are computed on the whole frequency gird, but only for the fiducial waveform. 

For our purposes, we have followed the implementation outlined in~\cite{Leslie:2021ssu} and the publicly available associated code~\cite{rel_code}. We have, however, used the waveform \texttt{IMRPhenomD\_NRTidalv2}~\cite{Dietrich:2019kaq} and employed the above code in conjunction with the sampling library \texttt{Bilby}, employing the code in~\cite{janquart:2022}.

\section{Results}
\label{sec:results}

We consider three different sources (A,B,C) with parameters chosen following mainly the injection study performed by the LIGO-Virgo-KAGRA collaboration~\cite{LIGOScientific:2018hze} to mimic GW170817. The properties of the sources used for injections are listed in Tab.~\ref{tab:sources}, where $m_i$, $\chi_i$, and $\Lambda_i$, with $i\in \{1,2\}$ are, respectively, the mass, spin, and dimensionless tidal deformability of the component neutron star, while $\mathcal{M}_c$ is the chirp mass and $\tilde{\Lambda}$ the binary's mass-weighted tidal deformability:
\begin{equation}
\tilde{\Lambda} = \frac{16}{3} \frac{(m_1 + 12 m_2)m_1^4 \Lambda_1 + (m_2 + 12 m_1)m_2^4 \Lambda_2}{(m_1+m_2)^5}.
\label{lambdat}
\end{equation}

All the simulated signals are injected at a distance $d_L = 100 \, \rm Mpc$ with inclination $\iota = 0.4$, zero polarization angle, and at a sky location $\left( \alpha, \delta \right) = \left( 1.375, -1.211 \right)$.
The priors used for the analysis are reported in Tab.~\ref{tab:priors}, where $\delta \tilde{\Lambda}$ is defined as

\begin{eqnarray} 
\delta \tilde \Lambda &=& \frac{1}{2}\left[\sqrt{1-4\eta}(1-\frac{13272}{1319}\eta 
+ \frac{8944}{1319}\eta^2)(\Lambda_1+\Lambda_2) \right. \nonumber\\ 
&& \,\,\,\,\,\,\,+ \left.(1-\frac{15910}{1319}\eta
+\frac{32850}{1319}\eta^2+\frac{3380}{1319}\eta^3)(\Lambda_1-\Lambda_2)\right], \nonumber\\
\end{eqnarray}
with $\eta=\frac{m_1m_2}{(m_1+m_2)^2}$ being the symmetric mass ratio of the binary.
\begin{table}[tbh]
	\setlength\extrarowheight{4pt}
	\begin{tabular}{ |c|c|c|c|c|c| }
		\hline
		
		Name	& $m_1$, $m_2$ & $\mathcal{M}_c$ & $\Lambda_1$, $\Lambda_2$ & $\tilde{\Lambda}$ & $\chi_1$, $\chi_2$  \\ [0.5ex]
		\hline
		Source A & 1.68, 1.13 & 1.19479 & 77, 973  & 303 & 0, 0\\
		Source B & 1.38, 1.37 &  1.19700 & 275, 309 & 292 & 0.02, 0.03 \\
		Source C & 1.38, 1.37 & 1.19700  & 1018, 1063  & 1040 & 0, 0 \\
		
		\hline
	\end{tabular}
	\caption{Source properties used for injections.}
	\label{tab:sources}
\end{table}

\begin{table}[tbh]
	\setlength\extrarowheight{4pt}
	\begin{tabular}{ |c|c| }
		\hline
		
		Parameter	& Range  \\ [0.5ex]
		\hline
		$\mathcal{M}_c$ & [$\mathcal{M}_{c,s} \pm 0.05$ ] \\
		q & [0.5, 1] \\
		$\chi_1$, $\chi_2$ & [0.0, 0.15] \\
		$d_L$ & [1,500] \\
		$\tilde{\Lambda}$ & [0,5000] \\
        $\delta \tilde{\Lambda}$ & [-5000, 5000] \\
		
		\hline
	\end{tabular}
	\caption{Priors employed in the PE analysis, where $\mathcal{M}_{c,s}$ represents the chirp mass injected value of the specific source analyzed. For the luminosity distance $d_L$, the prior is taken uniform in comoving volume; for all the other parameters, the prior is uniform in the indicated range.}
	\label{tab:priors}
\end{table}

\begin{figure}[t]
	\centering
	\includegraphics[width=1.\columnwidth]{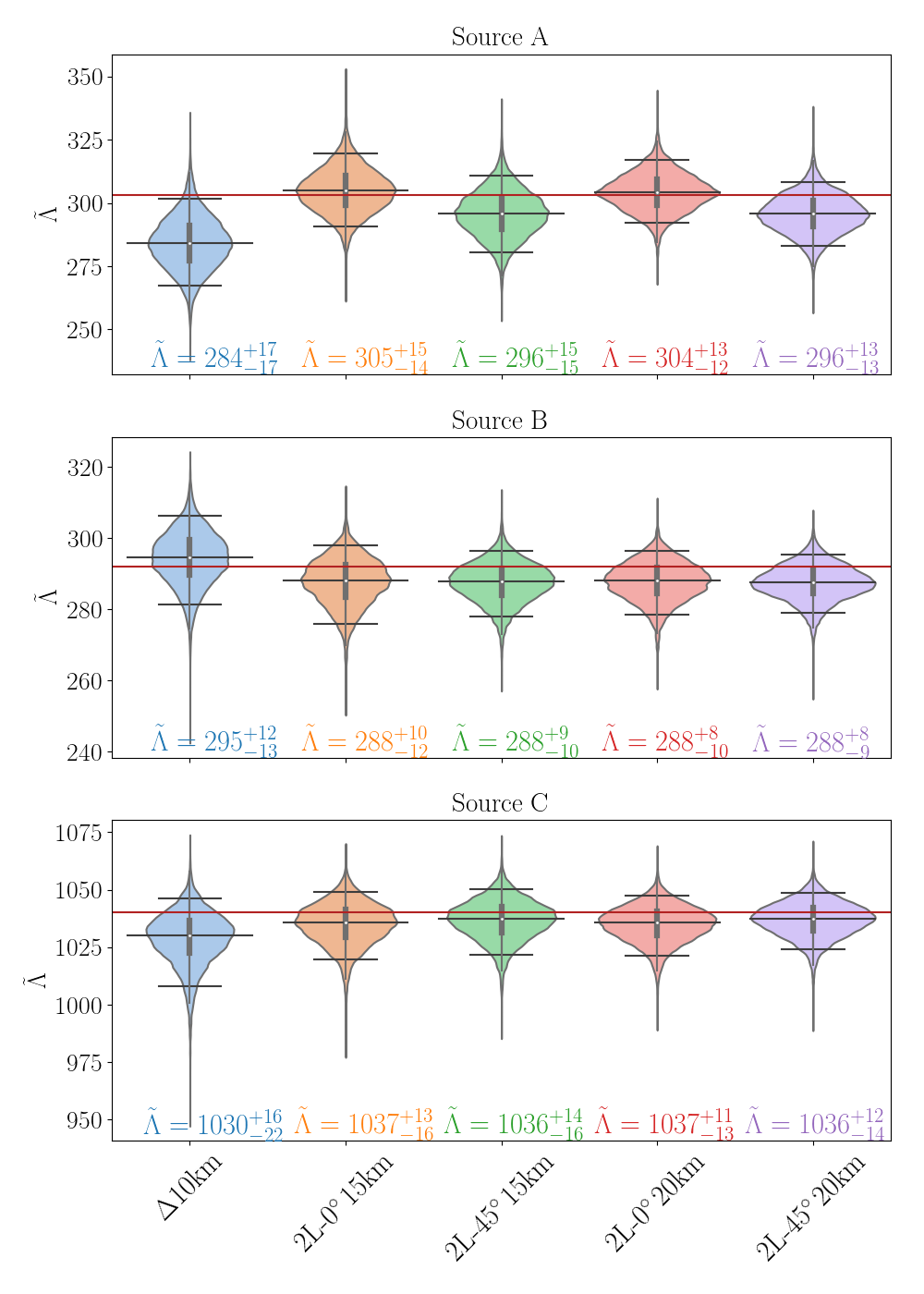} 
	\caption{Violin plots for the posterior density distribution of $\tilde{\Lambda}$ for the five reference detector configurations, and for all three sources in Table~\ref{tab:sources}: Source A (top panel), Source B (middle panel) and Source C (bottom panel). The black horizontal bars indicate the median value and the $90\%$ confidence interval, the black vertical lines mark the support of the posterior; the red horizontal line shows the injected value of $\tilde{\Lambda}$. For each posterior we also report the median, together with the $5\%-$ and $95\%-$ quantile values.}
	\label{fig:tot_sources_lambdas}
\end{figure}	

\begin{figure*}[t]
	\centering
	\includegraphics[width=1.\textwidth]{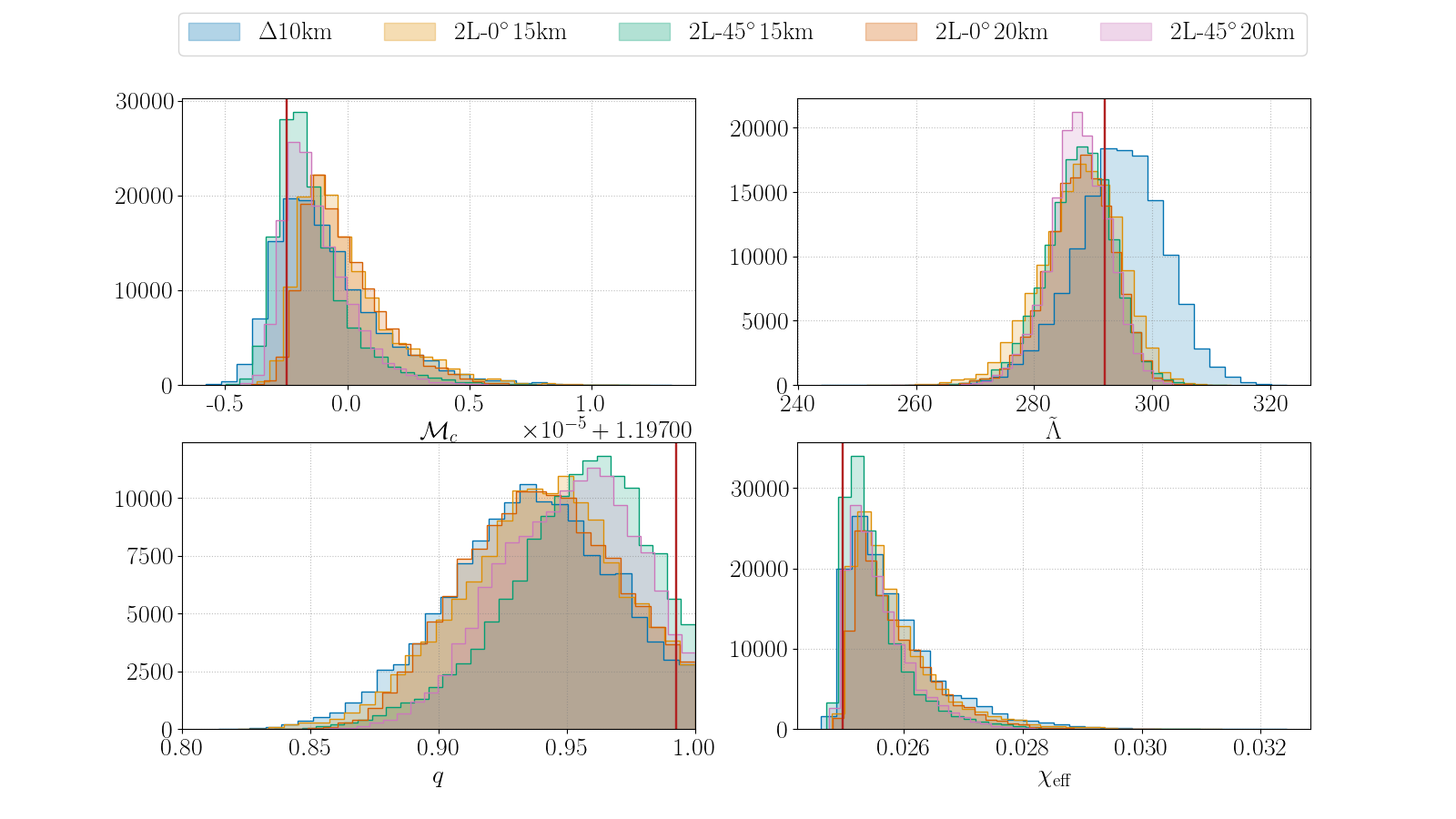} 
	\caption{Posteriors for chirp mass $\mathcal{M}_c$ (top left), tidal deformability $\tilde{\Lambda}$ (top right), mass ratio $q$ (bottom left), and effective spin $\chi_{\rm{eff}}$ (bottom right), for Source B, cf.~Tab.~\ref{tab:sources}. The different colors correspond to the various configurations considered, while the red line indicates the injected values.}
	\label{fig:totparams}
\end{figure*}

Two different shapes have been proposed for ET in Ref.~\cite{Branchesi:2023mws}: (i)~a single detector with a triangular configuration, henceforth called $\Delta$, (ii)~two L-shaped detectors in separate locations, with aligned or misaligned arms, henceforth called 2L-$0^{\circ}$ and 2L-$45^{\circ}$, respectively. The $\Delta$ configuration consists of three V-shaped detectors, with each V having a 60\degree opening angle between the arms. 
For 2L-$0^{\circ}$, the detectors have arms with the same orientation, while in the case of 2L-$45^{\circ}$ one detector has the arms rotated by 45\degree with the respect to the other one.

The $\Delta$ detector may have an arm-length of 10km or 15 km, while the L-shaped ones may have arm-lengths of 15km or 20km. As of now, there are two main candidate sites for ET \footnote{Recently, also a third location in Saxony, Germany, has been considered as a possible site option.}, one in Sardinia (Italy), and one in Limburg (at the border shared by Netherlands, Belgium and Germany). In this paper, the $\Delta$ detector is located in Sardinia, whereas in the case of two L-shaped detectors one is in Sardinia and the other one in Limburg. All detector configurations used in our work are listed in Table \ref{tab:config_table}.

One of the main challenges to reach the sensitivity planned for ET is dealing with quantum noise, which includes shot noise at high frequencies and radiation pressure noise at low frequencies. A high laser power reduces shot noise, but a low laser power is instead required to reduce radiation pressure noise. To counter this problem, ET will be practically composed of two interferometers working together in a xylophone configuration, one detector optimized for low frequencies and with a low laser power, and the other one optimized at high frequencies and with a high power. Low frequency sensitivity is also affected by thermal noise, therefore a further improvement is expected if the low frequency detector operates at cryogenic temperatures \cite{Hild:2010id}.
\newline In this section we present the results of PE runs, comparing the different detector configurations, arm-lengths and laser power. We also look at the improvement we get when analyzing data starting at different frequencies.

	\begin{center}
		\begin{table}[h]
			\setlength\extrarowheight{2pt}
			\begin{tabular}{ |c|c|c|c| }
				\hline
				
				Name	& Shape & Relative orientation & Arm-length  \\ [1ex]
				\hline
				$\Delta \rm 10km$ & Triangular & - & 10 km\\ [1ex]
				2L-$0^{\circ} \, \rm 15 km$ & 2 L-shaped & aligned & 15 km \\ [1ex]
				2L-$45^{\circ} \, \rm 15 km$ & 2 L-shaped & misaligned & 15 km  \\ [1ex]
				2L-$0^{\circ} \, \rm 20 km$ & 2 L-shaped & aligned & 20 km \\ [1ex]
				2L-$45^{\circ} \, \rm 20 km$ & 2 L-shaped & misaligned & 20 km \\ [1ex]
				$\Delta \rm 15km$ & Triangular & - & 15 km \\ [1ex]
				2L-$0^{\circ} \, \rm 10 km $ & 2 L-shaped & aligned & 10 km  \\ [1ex]
				2L-$45^{\circ} \, \rm 10 km$ & 2 L-shaped & misaligned & 10 km \\ [1ex]
				\hline
			\end{tabular}
			\caption{Names used throughout this work for the different detector configurations and arm-lengths.}
			\label{tab:config_table}
		\end{table}
	\end{center}

\begin{figure*}[t]
	\centering
	\includegraphics[width=1.0\textwidth]{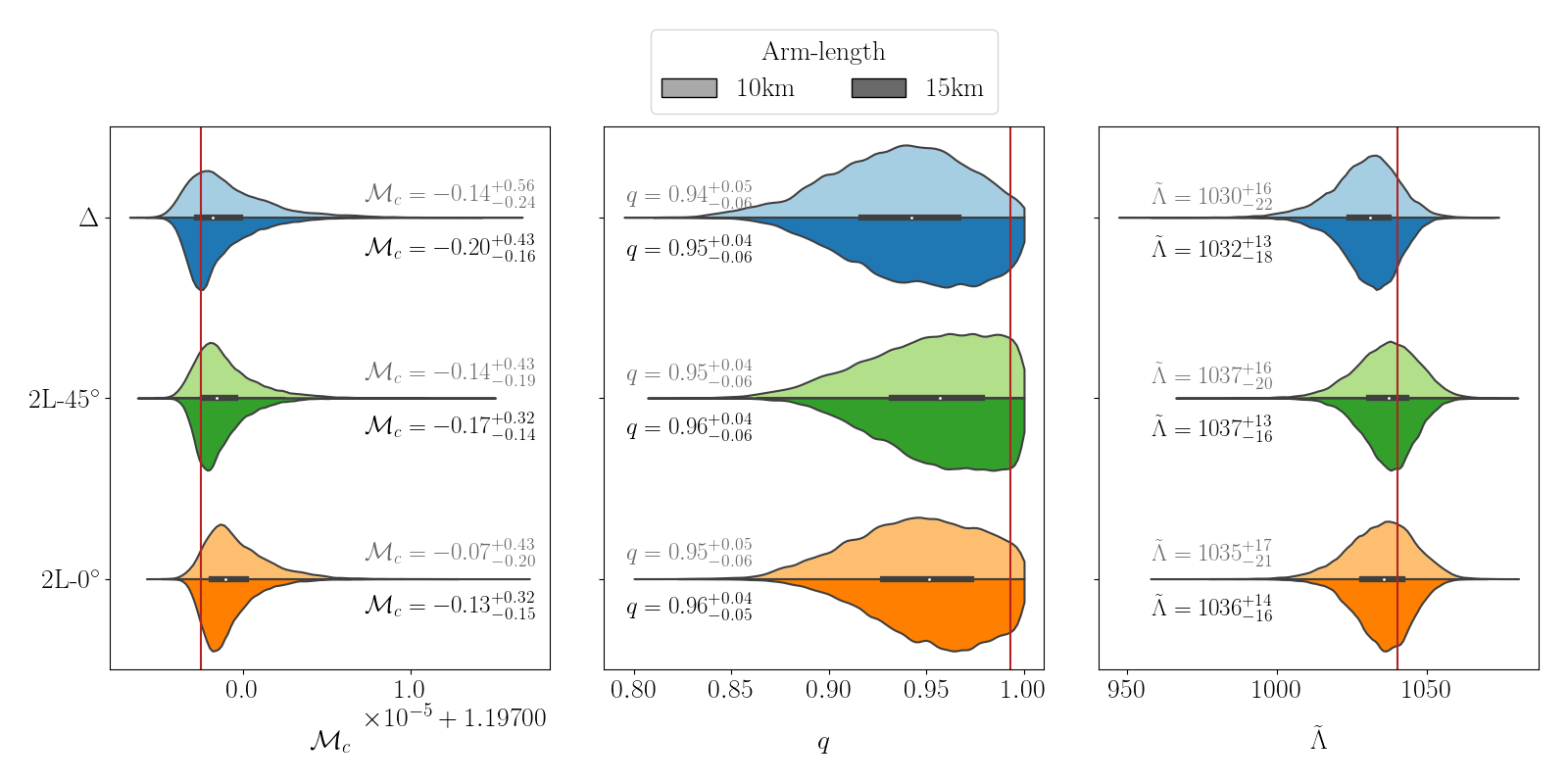} 
	\caption{Comparison of $\mathcal{M}_c$, $\tilde{\Lambda}$, and $q$ posteriors between $\Delta$, 2L-$0^{\circ}$ and 2L-$45^{\circ}$ with same arm-length, for Source C, cf.~Tab.~\ref{tab:sources}. The darker color shade refers to detectors with 15km arm-length, the lighter one to detectors with 10km arm-length. The red line indicates the parameter's injected value. For each parameter and configuration, we also report the median and $5\%$- and $95\%$- quantile values; for $\mathcal{M}_c$, also the reported values have an offset $\times 10^{-5} + 1.19700$.}
	\label{fig:source4_same_arm_violin}
\end{figure*}

\begin{figure*}[t]
	\centering
	\includegraphics[width=\textwidth]{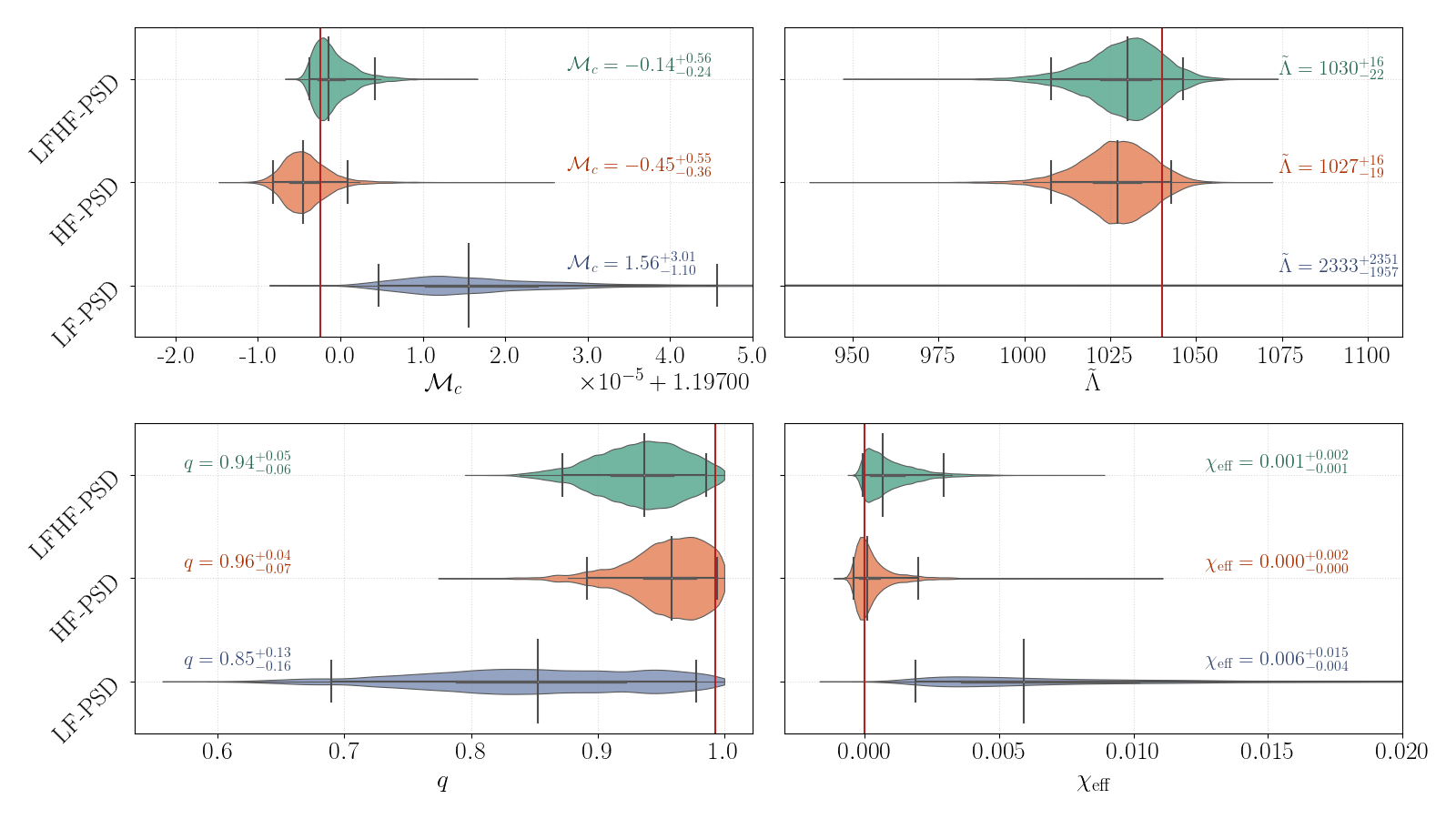} 
	\caption{Posteriors of $\mathcal{M}_c$, $\tilde{\Lambda}$, $q$ and $\chi_{\rm{eff}}$, for Source C (Tab.~\ref{tab:sources}) and the $\Delta \rm 10km$ configuration, recovered with the different PSDs.  The vertical black lines show the median and the 90\% confidence interval, while the black horizontal line indicates the support of the posterior; the red line shows the paramter's injected value. For $\mathcal{M}_c$, the reported median and quantile values have an offset $\times 10^{-5} + 1.19700$.}
	\label{fig:psd_tot}
\end{figure*}

\subsection{Detector configuration comparison}
s
First, we test how the different detector configurations ($\Delta$, 2L-$0^{\circ}$ or 2L-$45^{\circ}$) perform in PE analyses. We take into account only the five reference configurations~\cite{Branchesi:2023mws}: $\Delta$ with 10km arms, 2L-$0^{\circ}$ and 2L-$45^{\circ}$, both with 15km or 20km arms. For this comparison, all the runs are performed with starting frequency $f_{\rm{low}}$ = 10 Hz, and using the PSD curve for the xylophone configuration with the low frequency detector operating at cryogenic temperatures (`LFHF'). Figure~\ref{fig:tot_sources_lambdas} shows the posteriors for the tidal deformability parameter $\tilde{\Lambda}$ for the three different sources, reporting the median and $90 \%$ intervals for each configuration.   

For Source B, in Fig.~\ref{fig:totparams} we show also the posterior distributions for the other binary parameters: chirp mass $\mathcal{M}_c$, mass ratio $q$ and effective spin $\chi_{\rm{eff}}$, defined as 

\begin{equation}
\chi_{\rm{eff}} = \frac{\chi_1 m_1 + \chi_2 m_2}{m_1 + m_2},
\end{equation}

with $\chi_i$ and $m_i$ being the spin and mass of the component neutron star.
From Fig.~\ref{fig:totparams} and especially from Fig.~\ref{fig:tot_sources_lambdas} we clearly see an improvement going from the $\Delta$ to the two L-shaped detectors, since, for example, the width of the $90 \%$ interval on $\tilde{\Lambda}$ reduces between 12\% and 24\% when going from the $\Delta \rm 10km$ to the 2L-$45^{\circ}$ configuration.
However, we must take into account the fact that the $\Delta$ detector here is assumed to have a shorter arm-length, which has a great influence on the PSD, cf.~Fig.~\ref{fig:et_config}. For this reason, we also compare the $\Delta$ and 2L configurations assuming they have the same arm-length. Fig.~\ref{fig:source4_same_arm_violin} shows Source C posteriors for $\mathcal{M}_c$, $q$, and $\tilde{\Lambda}$, for the different configurations, but assuming the same arm-length. In this case, we do not see a strong difference in the parameters recovery, as indicated by the $5\%$ and $95\%$ quantile values reported in the plot. This suggests that the specific configuration does not have a major impact on the precision of parameter estimation. However, if the configuration choice is bound to a certain arm-length, e.g., due to limitations to the overall budget, one must take into account the improvements obtained with longer arms.

\subsection{Effect of varying PSDs}

\begin{figure}[t]
	\includegraphics[width=1\columnwidth]{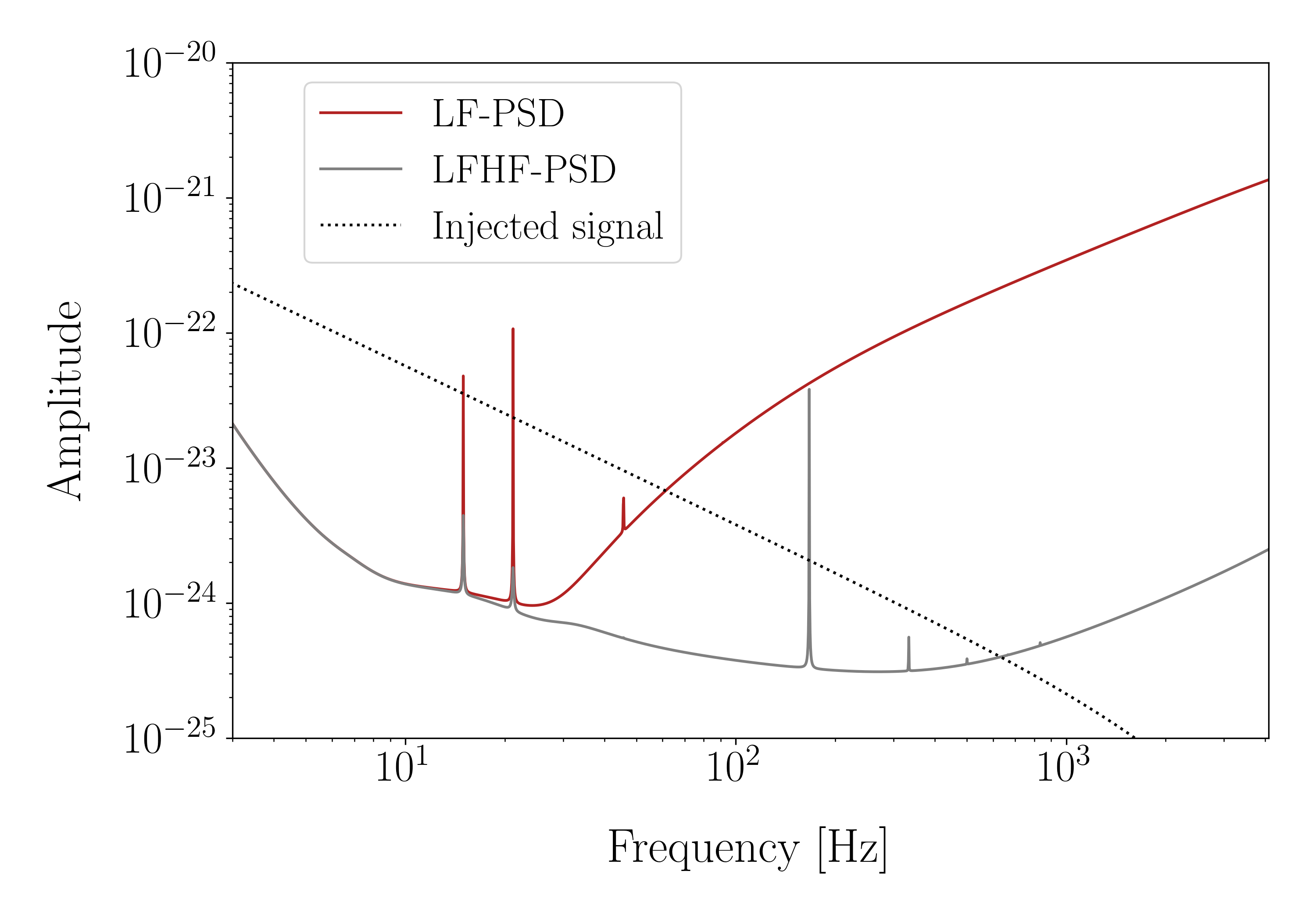} 
	\caption{Comparison between the PSD optimized at low frequencies (LF-PSD), in red, and the one of the xylophone configuration (LFHF-PSD), in grey, with a typical injected GW siganl (dotted line). Both PSDs refer to a detector with 10km arm-length. The sensitivity of the LF-PSD becomes too low at high frequencies to allow to detect the signal above $\sim 100 \rm{Hz}$.}
	\label{fig:psd_hf_signal}
\end{figure}

As mentioned in the previous sections, the current plan for ET includes a `xylophone' configuration, in which each detector is effectively composed of two interferometers, operating a high or low power laser. The high-power laser is expected to improve sensitivity at high frequencies, the low-power one, on the other hand, improves sensitivity below 30 Hz \cite{Hild:2009ns}\cite{Hild:2010id}. 
We perform the same PE analysis using the PSD of the different interferometers and we compare results. In particular, we study the PSD for the detector optimized at high frequencies (HF-PSD), the one optimized at low frequencies (LF-PSD), and the xylophone combination (LFHF-PSD), with the low-frequency interferometer operating at cryogenic temperatures. 
Since we are interested in the PSD's effect only, here we study just one source, Source B, and one detector configuration, $\Delta \rm 10km $, performing the analysis from a starting frequency $f_{\rm{low}}=10 \rm{Hz}$.
The posteriors for $\mathcal{M}_c$, $q$, $\tilde{\Lambda}$, and $\chi_{\rm{eff}}$ are shown in Fig \ref{fig:psd_tot}, where we also report the median and the $5\%$ and $95\%$ quantiles for each parameter. The PSD optimized at low frequencies performs much worse than the other ones, with a 90\% confidence interval 2.5 times larger in the case of mass ratio. $\tilde{\Lambda}$ is not recovered with the LF-PSD, while it is constrained with an accuracy of almost 4\% in the other cases. $\tilde{\Lambda}$ represents an extreme case, since its contribution enters the gravitational-wave phase mainly at high frequencies, from a few hundreds Hz and above \cite{Dietrich:2020eud,Harry:2018hke}, and therefore is affected by the shape of LF-PSD, as shown in Fig \ref{fig:psd_hf_signal}, more than other parameters. In general, we obtain a much worse parameter recovery when using the LF-PSD alone, meaning that, if the preferred solution of a xylophone implementation is not available, the high-frequency optimized PSD is favorable, in particular if we want to constrain $\tilde{\Lambda}$. 

\subsection{Effect of varying minimum frequency}

\begin{figure*}[t]
	\includegraphics[width=1\textwidth]{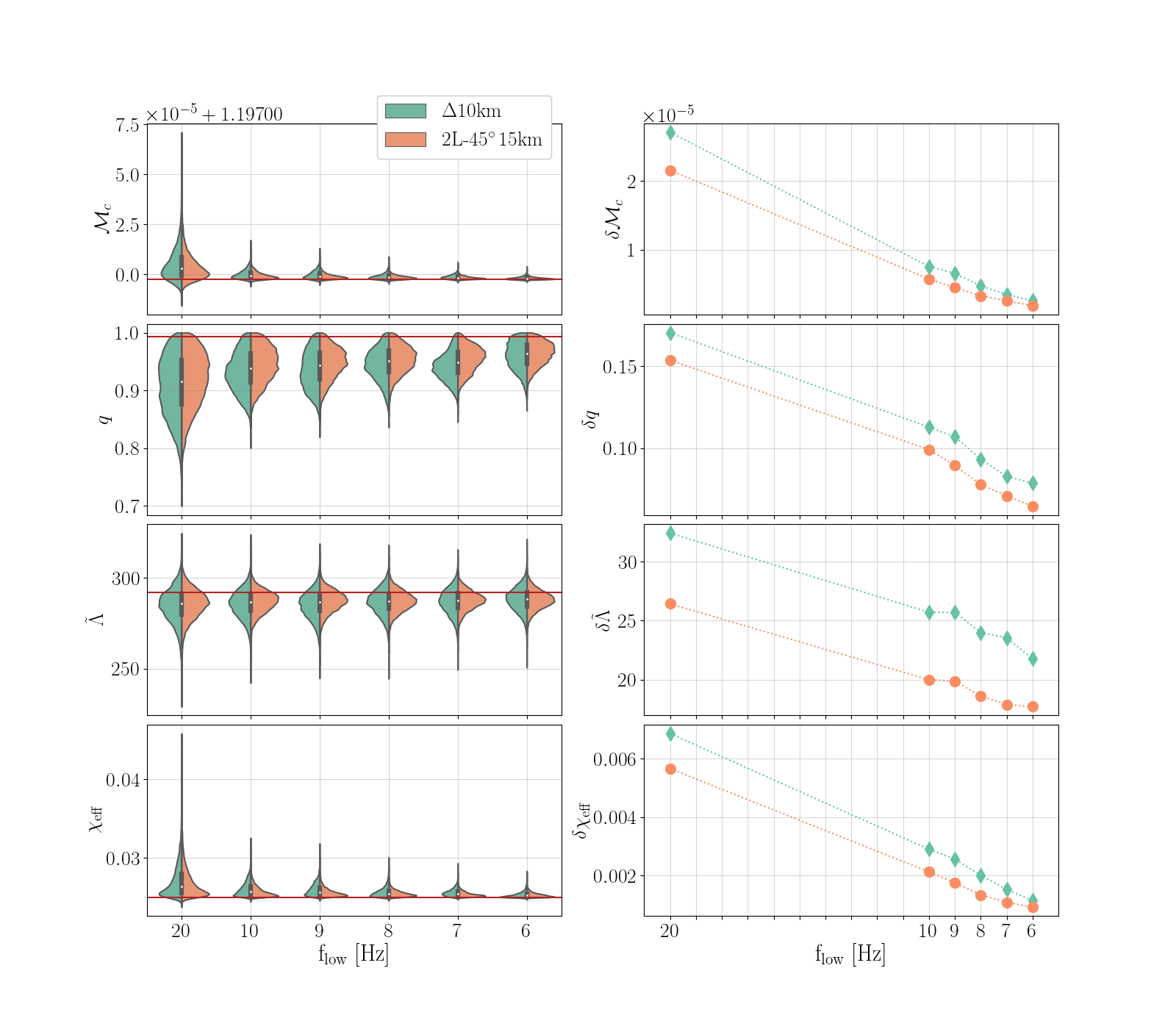} 
	\caption{\textit{Left:} Posterior distributions for $\mathcal{M}_c$, $q$, $\tilde{\Lambda}$, and $\chi_{\rm eff}$ for the different choices of $f_{\rm{low}}$ and for Source B (Tab.~\ref{tab:sources}), in green for the $\Delta$ configuration with 10km arms, in orange for the 2L-$0^{\circ}$ with 15km arms. The red horizontal lines correspond to the parameters' injected values. \textit{Right:} Width of the 90\% confidence interval for each parameter, as a function of the value of $f_{\rm{low}}$. }
	\label{fig:flow_comp}
\end{figure*}

A big achievement of the ET is to improve the sensitivity at low frequencies. Therefore, in this section, we will study the impact of choosing different starting frequencies in the PE analysis. We note that lowering the starting frequency by only a few Hz has a huge impact on the duration of the waveform, and therefore of the computational cost of the analysis. 
We analyze injections with parameters of Source B, for two different configurations, $\Delta \rm 10km$, and 2L-$45^{\circ} \, \rm 15km$. In each case, we perform PE tests with the following starting frequencies: 6, 7, 8, 9, 10, and 20 Hz. In this case, the analysis is performed in zero noise, since we want to focus on the impact of $f_{\rm{low}}$ without risking to take into account possible fluctuations induced by the noise realizations. In fact, specific noise realizations can cause a shift in the posterior recovered for $\tilde{\Lambda}$. This shift is usually within 5\% of the actual $\tilde{\Lambda}$ value, and, therefore, goes unnoticed in the analysis performed with current detectors. However, we showed that for ET the precision with which $\tilde{\Lambda}$ can be measured improves noticeably. This means that, due to these shifts, we might end up seeing the injected values lying outside the posterior's 90\% interval when simulations are done in Gaussian noise. When real ET data is being analyzed later, this is a point that must be evaluated very carefully. For our purposes, up to now, we compared only results obtained with the same $f_{\rm{low}}$, and as long as we use the same noise seed, we expect a possible fluctuation to affect all the runs in the same way, and, therefore, to be not relevant in our comparison. Here, however, we use different starting frequency, leading to longer signals and more cycles being analyzed. In this case, the noise fluctuation's outcome can be different for the different $f_{\rm{low}}$ used. To quantify this, in Table~\ref{tab:noise_flow} we report the median and 90\% interval values of the posteriors on $\tilde{\Lambda}$, obtained from analysis both in gaussian and zero noise, with the same seed but different $f_{\rm{low}}$. While we see fluctuations in the median values recovered from gaussian noise runs, in the zero noise case the median is almost constant. Therefore, to compare $\tilde{\Lambda}$ recovery with different starting frequency, we look at zero noise injections. Fig.~\ref{fig:flow_comp} shows the posteriors and their 90\% width for the different $f_{\rm{low}}$ values. We see a clear improvement when going to lower frequencies, especially for the recovery of chirp mass. The plot also highlights how in general the 2L-$45^{\circ} \, \rm 15km$ configuration yields tighter constraints on the parameters' posteriors, but we stress again that the main impact is given by the arm-length, not the configuration per se.

\begin{table}[!htb]
	\setlength\extrarowheight{4pt}
	\begin{tabular}{|c|c|c|c|}
		\hline
		&      $f_{\mathrm{low}}$ & Gaussian noise & zero noise  \\[1ex] \hline
		2L-$45^{\circ} \, \rm 15km$ & 7 Hz & $279.31^{+ 9.55}_{- 11.35}$ & $287.90^{+ 8.35}_{-9.54}$ \\[1ex] \hline
		& 8 Hz & $292.90_{-8.33}^{+8.40}$ & $287.36^{+9.01}_{-9.59}$  \\[1ex] \hline
		& 9 Hz & $278.63_{10.72}^{9.75}$     & $287.65 _{-10.92}^{+8.95}$   \\[1ex] \hline
		& 10 Hz & $285.74_{-13.26}^{11.15}$ & $287.40_{-10.96}^{+9.03}$ \\[1ex] \hline
		& 20 Hz & $296.72_{-11.66}^{+10.46}$ & $286.18^{+12.01}_{-14.42}$ \\[1ex] \hline
		$\Delta \rm 10 km $    & 7 Hz & $277.27_{-13.52}^{+11.87}$    & $287.29_{-12.59}^{+10.95}$   \\[1ex] \hline
		& 8 Hz & $292.95_{-9.77}^{+9.15}$ & $287.11_{-12.79}^{+11.21}$ \\[1ex] \hline
		& 9 Hz & $274.62_{-14.09}^{+12.78}$    & $285.51_{-13.97}^{+11.71}$   \\[1ex] \hline
		& 10 Hz & $273.16_{-17.09}^{+15.18}$ & $285.51_{-13.97}^{11.71}$ \\[1ex] \hline
		& 20 Hz & $280.09_{-17.10}^{+13.76}$ & $285.21^{+14.56}_{-17.86}$ \\[1ex] \hline
	\end{tabular}
\caption{Recovered $\tilde{\Lambda}$ values with $90 \%$ intervals, comparing Gaussian-noise and zero-noise runs for different values of $f_{\rm low}$.}
\label{tab:noise_flow}
\end{table}

\section{Summary}
\label{sec:conclusion}

We performed PE studies to compare the different proposed designs for ET. We focus on BNS systems, in particular, to find out how well the tidal effects will be measured. We compare different detector shapes, considering a single triangular detector and two L-shaped ones. In the latter case, we investigate both the cases of aligned or misaligned detectors. Moreover, ET will be composed of two interferometers, one optimized at high and one at low frequencies. We compared results obtained using the different PSDs, the low- and high- frequency one, as well as the xylophone PSD, obtained by combining the two. Finally, we looked at how the PE results improve when using lower cutoff frequencies, investigating $f_{\rm low} = 20,10,9,8,7,6$ Hz. We find that:

\begin{itemize}
	
	\item The shape and alignment of the detectors have very little influence on the recovery of parameters.\footnote{However, we want to point out that we have not include information from the null stream.}
	
	\item The chosen arm-length, instead, plays an important role, as expected given its effect on the PSD. This means that, when comparing the currently proposed configurations, the $\Delta \rm 10km$ one performs worse, but this is merely due to the fact that it has a shorter arm-length with respect to the 2L ones. When comparing the different configurations, and assuming the same arm-length, we find no significant difference in the results.
	
	\item The constraints recovered with the LF-PSD are much worse than the other ones, especially with respect to $\tilde{\Lambda}$. This is expected since with the LF-PSD the signal above a hundred Hz is not detectable.
	
	\item Noise fluctuations have a very strong impact on the $\tilde{\Lambda}$ measurement, causing the posteriors' median values to shift. With ET, $\tilde{\Lambda}$ will be measured with a very high accuracy, therefore, although such shifts are of the order of a few percent, they can be enough for the injected value to lie outside the support of the posterior.
	
	\item Regarding the different cutoff frequencies, we studied two different detector configurations, $\Delta \rm 10km$ and 2L-$45^{\circ} \, \rm 15km$, and find no substantial difference between them. The parameters posteriors become clearly tighter when going to lower frequencies. This is particularly evident in the case of $\mathcal{M}_c$ posteriors, but an improvement is also present for $\tilde{\Lambda}$, but only of about $20\%$.
\end{itemize}

\begin{acknowledgments} 

A.P. ~is supported by the research programme of the Netherlands Organisation for Scientific Research (NWO).
This work was performed using the Computing Infrastructure of Nikhef, which is part of the research program of the Foundation for Nederlandse Wetenschappelijk Onderzoek Instituten (NWO-I), which is part of the Dutch Research Council (NWO). 
A.S. thanks the Alexander von Humboldt foundation in Germany for a Humboldt fellowship for postdoctoral researchers. 
The authors are grateful for computational resources provided by the LIGO Laboratory and supported by the National Science Foundation Grants No.~PHY-0757058 and No.~PHY-0823459. 
Particularly, we thank Michael Thomas for prompt help with 
computing issues. 
This research has made use of data, software and/or web tools obtained from the Gravitational Wave Open Science Center (https://www.gw-openscience.org), a service of LIGO Laboratory, the 
LIGO Scientific Collaboration and the Virgo Collaboration. LIGO is funded by 
the U.S. National Science Foundation. Virgo is funded by the French Centre 
National de Recherche Scientifique (CNRS), the Italian Istituto Nazionale della 
Fisica Nucleare (INFN) and the Dutch Nikhef, with contributions by Polish and Hungarian institutes.
  
\end{acknowledgments}

\bibliography{refs}

\end{document}